# PROTECTION OF THE 13 T NB3SN FRESCA II DIPOLE

Ph. Fazilleau, M. Devaux, M. Durante, T. Lecrevisse, J.-M. Rey, CEA DSM-IRFU-SACM, Gif-sur-Yvette, France

*Abstract*

The EuCARD project aims on construction of a 19 T hybrid dipole; it will be made of a 6 T HTS dipole associated to a 13 T outsert $Nb_3Sn$ dipole [1]. This paper reviews the quench analysis and protection of the 13 T $Nb_3Sn$ dipole.

## INTRODUCTION

To study the protection of the dipole, first computations have been made with the QTRANSIT fortran code, which is a 3D simulation of the quench thermal transient in the magnet, based on the quench propagation velocities and the resistance growth with time. Even if it is a very reliable code, used and compared with test results on several magnets, it has been mainly applied to large magnets, indirectly cooled. Moreover, the usual propagation velocities formulas depend strongly on the magnetic field (magneto-resistance, current sharing temperature) and they are obviously not uniform within the winding. So we decided to further study the protection with the Finite Element Model (FEM) Cast3M [2], with modified procedures to take into account the decrease of the current with time, directly related to the joule losses dissipated within the winding.

The ignition and expansion of the quench has been studied with 3D models in the low field zone as well as in the high field zone: longitudinal propagation velocities have been computed and the internal resistive voltage gives the time needed to exceed the detection threshold. Once the detection is done and validated, the main contactor opens, with a discharge of the magnet into the dump resistor; heaters are activated and the FEM problem is reduced to 2D computations as the heaters are located all along the dipole.

The dipole is made of four double-pancakes: the coils of Fresca II are wound with a Rutherford cable composed of 40 strands with a diameter of 1 mm, with *Cu/Sc* = 1:3. Its length is 1.5 m and its total energy is 5.4 MJ. With a cold mass of 236 kg, its energy density is 18.4 J/g, comparable to the other magnets of same type [3].

## PROTECTION PRINCIPLE

To start with the protection of a magnet, we can calculate analytically simple figures: if we consider all the energy dissipated uniformly in the dipole, the mean temperature is 126 K. If it is only in one pole, the mean temperature is 182 K. In a quarter of the magnet, the mean temperature jumps to 276 K. These figures, easily calculable, leads to two obvious conclusions: we need heaters in order to spread the quench within the largest volume of the dipole and the detection must be as fast as possible.

Consequently, the protection principle is the extraction of the energy into a dump resistor as well as the growth of the internal resistance due to heaters. Figure 1 is a sketch of the electrical circuit; the value of the external resistor is 95 mΩ, set so that the voltage at the terminals of the magnet never exceeds 1 kV, ±500 V to ground by means of the grounding circuit. The resistance volume has also been set so that the voltage at its terminals remains maximal as long as possible. This leads to a total volume of 2.63 liters.

Heaters are located on the sides of the double pancakes: two or four could be used by pole and computations will guide us to the correct choice. We set their power is 50 W/cm$^2$ and they will cover 50% of the total allowable surface.

The Fresca II dipole will operate with a YBCO insert magnet of 6 T; the latter is not taken into account in the protection study as its energy and self-inductance are low compared to those of the dipole. Consequently the discharge of the current is driven by the main $Nb_3Sn$ magnet [4].

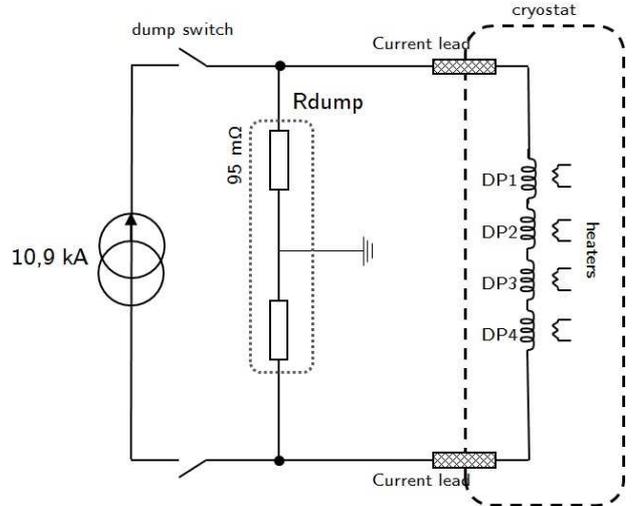

Figure 1: Protection circuit

## AFTER THE DETECTION: 2D COMPUTATIONS

As detailed supra, once the quench has been detected, the heaters are activated and the FEM problem is reduced to 2D computations. We are able to compute the temperature of the hot spot within the dipole as well as the current decrease evolution.

$$\int_{T_0}^{T_{max}} \frac{C_p(T)}{\rho(T)} dT = \int_0^\infty j^2(t)dt = \int_0^{t_{det}} J_0^2(t)dt + \int_{t_{det}}^\infty j^2(t)dt \quad (1)$$

This evolution is of primary importance as we can simply calculate the maximal temperature by applying the adiabatic hot spot criteria (2) taking into account the nominal current during a time $t_{det}$ followed by a fast discharge as shown in Figure 2.

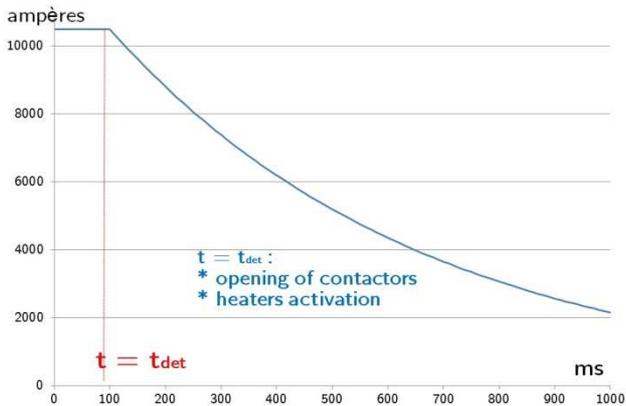

Figure 2: Current evolution

*Results of computations*

Figure 3 shows the evolution of the volume of the resistive zone: it takes 20 ms after the activation of the heaters for quenching the first conductor, located in the highest field zone, where the margins are reduced. This delay is due to the thermal barrier made of kapton and insulation, located between heaters and conductors. The dipole is totally resistive after 430 ms.

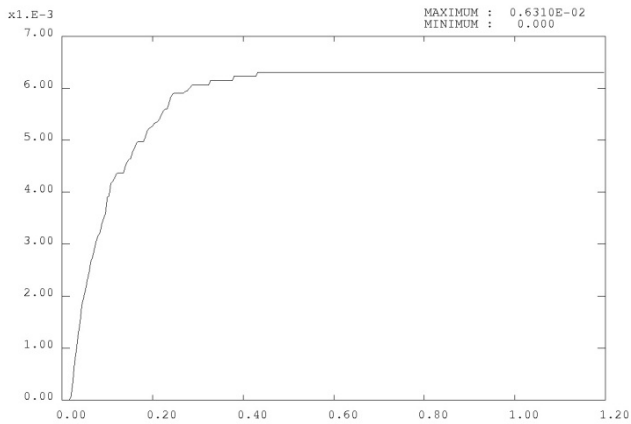

Figure 3: Evolution of the resistive zone volume

The current decrease evolution is shown in Figure 4. The time constant ($I/e$) is 0.57 s and it takes approximately 1.4 s to totally discharge the magnet. This evolution is used to calculate the hot spot temperature.

The dump resistor value increases by 30% during the discharge. Its temperature increases and reaches 638 K at the end of the discharge, which is an acceptable value. The voltage at its terminals, represented in Figure 5, remains maximal at the beginning of the discharge (nul slope) as we used for computations the volume mentioned *supra*; 65% of the total magnetic energy stored is dissipated in the dump resistor. The temperature field within the magnet at the end of the discharge is given in Figure 6.

*Adiabatic hot spot criteria*

We have to check that the current evolution has been correctly computed before using it to calculate the adiabatic hot spot criteria taking into account the detection time $t_{det}$. So we compare the maximal temperature evolution from the 2D computation results with the hot spot temperature calculated with the current evolution. The evolutions are remarkably close to each other as shown in Figure 7.

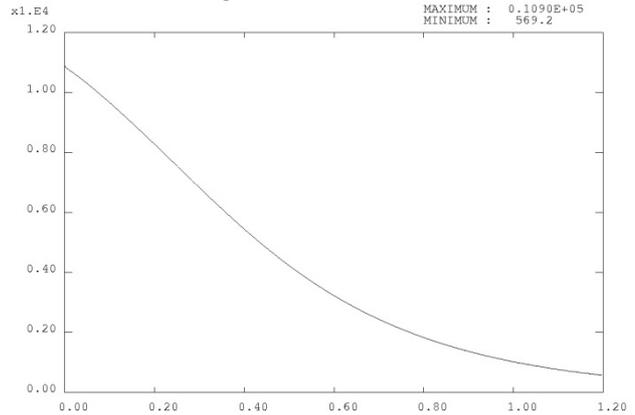

Figure 4: Evolution of the current

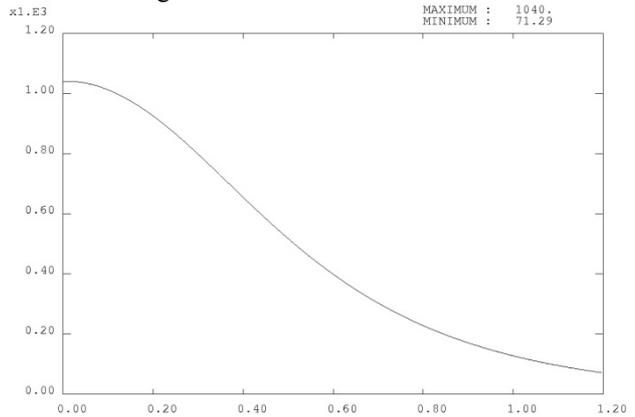

Figure 5: Evolution of the resistive voltage

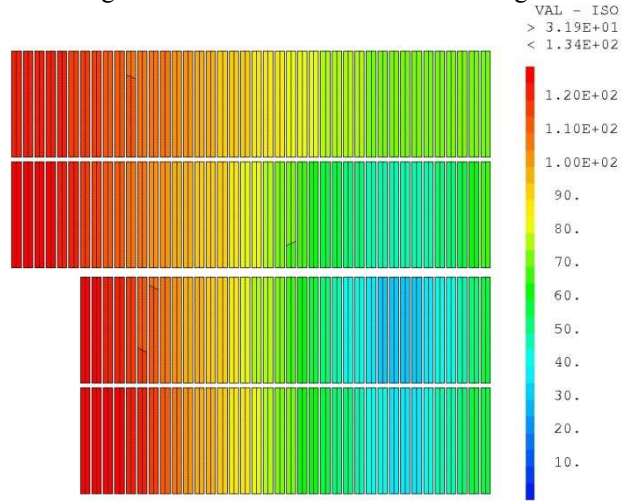

Figure 6: Temperature field at the end of the discharge

We can now calculate confidently the hot spot temperature taking into account the detection time $t_{det}$ and the results are shown in Figure 8. The detection must be lower than 25 ms if we want a maximal temperature below 150 K (four heaters case). With a detection time of 100 ms, the maximal temperature is 220 K for four heaters. The maximal temperature difference is around 30 K between two and four heaters.

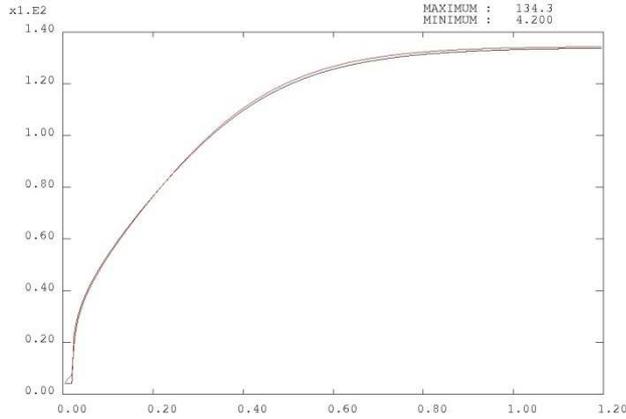

Figure 7: Comparison of the hotspot temperature evolution between adiabatic calculation and FEM computations

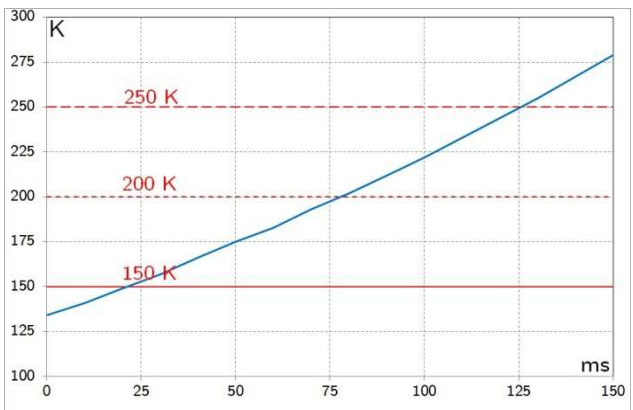

Figure 8: Hot spot temperature vs detection time

The 2D study demonstrates that the magnet is not endangered: the use of four heaters decreases the maximal temperature - with value around 150 K to 250 K depending on the detection time - in the dipole and helps to distribute more uniformly the temperature (lower temperatures gradients).

## BEFORE THE DETECTION: 3D COMPUTATIONS

After the 2D computations, we could have definitively concluded that the requested detection times are usually in the range 10 to 100 ms; nevertheless, we have been studying the propagation of the quench in the dipole with a 3D model. This is of great interest especially in the low field region as the quench propagates very slowly. The propagation is mainly longitudinal but the transverse one is also taken into account - especially in the high field region where the quench velocity is higher.

### MPZ: benchmark of Cast3M

Before studying in details the propagation in the whole magnet, we have studied the minimum propagating zone in a single conductor to benchmark the FEM Cast3M code.

The heat equation for the 1D static case without helium cooling leads to the minimum propagating zone (MPZ) formula stated in equation (3).

$$l_{MPZ} = \pi \sqrt{\frac{\lambda(T_c - T_{cs})}{\rho j^2}} \quad (2)$$

In the low field region, $l_{MPZ}$ = 26.5 mm.
In the high field region, $l_{MPZ}$ = 5 mm.

We inject in a unitary volume a pulse of energy and increase it up to the limit between recovery and expansion of the quench; the length of the resistive zone is the MPZ computed via the 3D model of the conductor. The injected energy at the limit should be the minimum quench energy (MQE). Nevertheless, it is strongly dependent on the step time used in the computations, contrary to the MPZ which is far less dependent. That is the main reason why we have studied the MPZ instead of the MQE.

Figures 9 and 10 show the results of the computations. The MPZ in the high field zone is 4.45 mm, a little bit lower than 5 mm, calculated with the formula (3). The agreement is even better in the low field region as the results of computations and formula calculation give the same value of 26.5 mm.

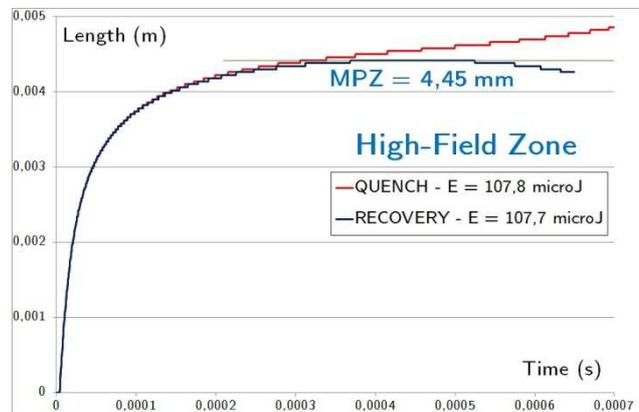

Figure 9: Length of the resistive zone - high field region

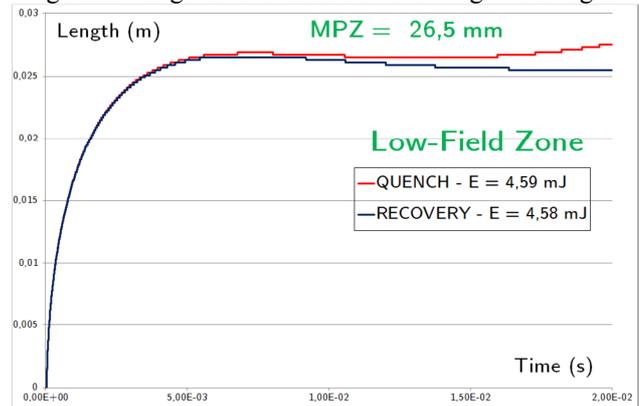

Figure 10: Length of the resistive zone - low field region

The 3D model has been successfully benchmarked and if we quench a zone longer than the computed MPZ, we will initiate a quench that will spread in the whole magnet.

*Propagation in the low field region*

We now study the propagation of a quench in a 3D model of the complete dipole. After some transients due to the energy pulse, the propagation velocity in the low field zone (0.5 T averaged on a single conductor) tends to the value of 0.6 m/s. As above stated, in this region the quench expands very slowly in the magnet.

The internal voltage of the resistive zone is shown in Figure 11. It takes 361 ms to exceed 100 mV. Nevertheless, due to the material characteristics in low field-especially the magnetoresistance-, the heating of the conductor is also very slow and the maximal temperature is only 47 K, 361 ms after the ignition of the quench, as shown in Figure 12.

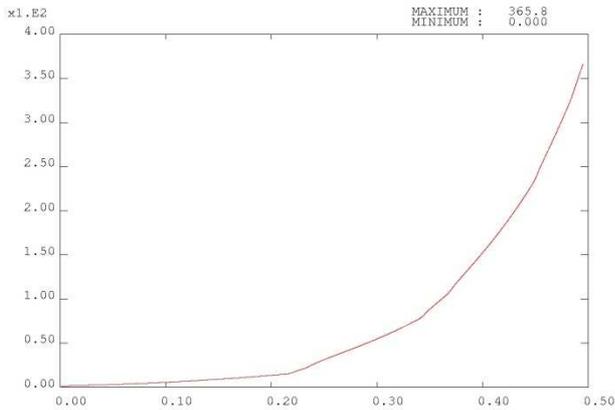

Figure 11: Resistive internal voltage in the low field region

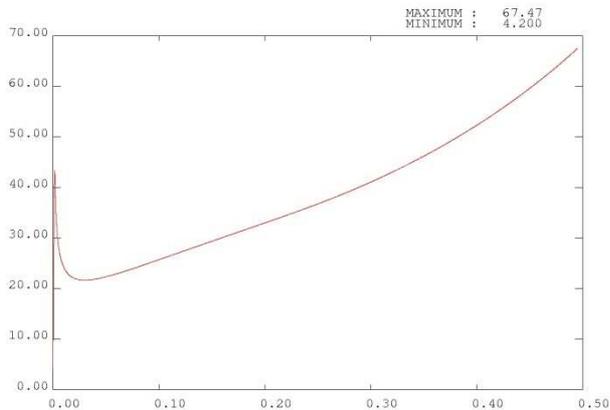

Figure 12: Maximal temperature in the low field region

We can then compute the hot spot temperature from this point, adding 10 ms for the validation of the quench, 5 ms for the switch opening, and then injecting the current evolution computed from the 2D computations. If a quench occurs in the low field zone, with a detection threshold of 100 mV, the maximal temperature is 135 K.

*Propagation in the high field region*

The propagation is obviously higher with a velocity of 6 m/s. Figure 13 shows that it takes 28 ms to exceed 100 mV. The maximal temperature within the conductor is 33 K as shown in Figure 14.

If a quench occurs in the high field zone, with a detection threshold of 100 mV, validation time of 10 ms, switch opening of 5 ms, the maximal temperature of the dipole at the end of the discharge is 157 K. We can compare this value with the 2D results taking into account a detection time of 43 ms; the temperature is slightly higher with a value of 166 K, as the adiabatic hot spot criteria is conservative.

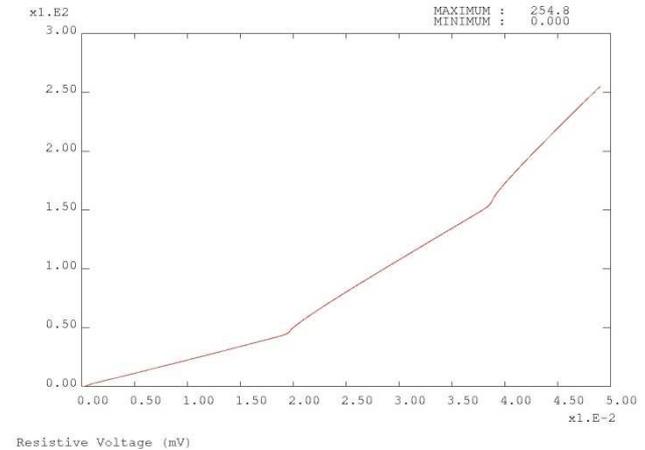

Figure 13: Resistive internal voltage in the high field region

## CONCLUSIONS

The quench study has been splitted in two parts. Before the detection, the quench ignition and expansion and the longitudinal propagation have been studied. The FEM based on Cast3m 3D has been benchmarked through estimation of the minimum propagating zone. The 3D computations in the low field zone show that the propagation is slow with a longitudinal velocity of 0.6 m/s but so is the temperature elevation; with a detection threshold of 100 mV, the maximal temperature is 135 K. In the high field region, the velocity is ten times larger and the 100 mV threshold is exceeded after 28 ms.

After the detection, heaters and dump resistor are taken into account and the problem, reduced to 2D computations, deals with the transverse propagation. Four heaters are needed to reduce thermal gradients. The voltage threshold of 100 mV led to a detection time of approximately 40 ms and a maximal temperature of approximately 160 K.

The Fresca II dipole will then not be endangered, mechanically or thermally, if the activation of the heaters is effective; the activation system must be redundant to avoid any fault scenario, invoking high thermal gradients and high temperature.

## ACKNOWLEDGEMENTS

This work was supported by the European Commission under the FP7 Research Infrastructures project EuCARD, and is part of EuCARD Work Package 7: Superconducting High Field Magnets (HFM) for higher luminosities and energies.